\documentclass[doublecol]{epl2} 
\usepackage{graphicx}
\usepackage{color}
\usepackage{epsfig}
\usepackage{amssymb}
\usepackage{amsfonts}
\usepackage{amsmath}

\usepackage{multicol}

\twocolumn

\title{Neutral bremsstrahlung electroluminescence in noble liquids}

\author{E. Borisova\inst{1,2} \and A. Buzulutskov\inst{1,2}\footnote{Corresponding author. Email: A.F.Buzulutskov@inp.nsk.su}}
\shortauthor{E. Borisova \etal}

\institute{
	\inst{1} Budker Institute of Nuclear Physics SB RAS, Novosibirsk, 630090, Russia\\
	\inst{2} Novosibirsk State University, Novosibirsk, 630090, Russia
}

\pacs{95.55.Vj}{Neutrino, muon,pion, and other elementary particle detectors; cosmic ray detectors} \pacs{61.25.Bi}{Liquid noble gases} \pacs{95.35.+d}{Dark matter}

\abstract{Proportional electroluminescence (EL) is the physical effect used in two-phase dark matter detectors, to optically record in the gas phase the ionization signal produced by particle scattering in the liquid phase. In our previous works the presence of a new EL mechanism in noble gases, namely that of neutral bremsstrahlung (NBrS), was demonstrated both theoretically and experimentally, in addition to the ordinary EL mechanism due to excimer emission. In this work we show that the similar theoretical approach can apply to noble liquids, namely to liquid helium, neon, argon, krypton and xenon. In particular, the photon yields and spectra for NBrS EL in noble liquids have for the first time been calculated, using the electron energy and transport parameters obtained in the framework of Cohen-Lekner and Atrazhev theory. The relevance of the results obtained to the development of noble liquid detectors for dark matter searches and neutrino experiments is discussed.}

\begin{document}
	
	\maketitle
	
	\section{Introduction}
	
	In noble liquid detectors for dark matter searches \cite{Chepel13} and low-energy neutrino experiments \cite{Majumdar21}, the scattered particle produces two types of signals: that of primary scintillation, produced in the liquid and recorded promptly ("S1"), and that of primary ionization, produced in the liquid  and recorded with a delay ("S2"). In two-phase (liquid-gas) detectors \cite{Akimov21}, to record the S2 signal proportional electroluminescence (EL) is used produced by drifting electrons in the gas phase under high enough electric fields. 
	
	According to modern concepts~\cite{Buzulutskov20}, there are three mechanisms responsible for proportional EL in noble gases: that of excimer (e.g. Ar$^*_2$) emission in the vacuum ultraviolet (VUV) \cite{Oliveira11}, that of emission due to atomic transitions in the near infrared (NIR) \cite{Oliveira13,Buzulutskov17}, and that of neutral bremsstrahlung (NBrS) emission in the UV, visible and NIR range \cite{Buzulutskov18}. These three mechanisms are referred to as excimer (ordinary) EL, atomic EL and NBrS EL, respectively. 
	
	NBrS EL is due to bremsstrahlung of drifting electrons scattered on neutral atoms:
	\begin{eqnarray}
		\label{Rea-NBrS-el}
		e^- + \mathrm{A} \rightarrow e^- + \mathrm{A} + h\nu \; . %\\
	\end{eqnarray}
	The presence of NBrS EL in two-phase Ar detectors has for the first time been demonstrated in our previous work~\cite{Buzulutskov18}, both theoretically and experimentally. Recently, the similar theoretical approach has been applied to all noble gases, i.e. overall to He, Ne, Ar, Kr and Xe, to calculate the photon yields and spectra for NBrS EL \cite{Borisova21}. NBrS EL in noble gases was  further studied experimentally in  \cite{Bondar20,Tanaka20,Kimura20,Takeda20,Takeda20a,Aoyama21,Aalseth21,Monteiro21} and theoretically in \cite{Amedo21}.
		
	On the other hand, much less is known about proportional EL in noble liquids \cite{Buzulutskov20,Masuda79,Schussler00,Aprile14,Ye14,Lightfoot09,Stewart10}. In a sense, the experimental data are even confusing. Indeed, in liquid Ar the observed threshold in the electric field for proportional EL, of about 60 kV/cm \cite{Buzulutskov20,Lightfoot09}, was 2 orders of magnitude less than expected for excimer EL \cite{Stewart10}. In liquid Xe, the EL threshold was more reasonable, around 400 kV/cm, but some puzzling EL events were observed below this threshold \cite{Aprile14}. 
	
	In our previous works \cite{Buzulutskov18,Buzulutskov20} it was suggested that these puzzling events at unexpectedly low fields might be induced by proportional EL produced by drifting electrons in a noble liquid due to NBrS effect, the latter having no threshold in the electric field. In this work we verify this hypothesis, namely we extend the theoretical approach developed for noble gases to noble liquids in order to develop a quantitative theory that can predict the photon yields and spectra for NBrS EL in all noble liquids. 
	
	What is new in this work is that the electron energy and transport parameters in noble liquids are calculated in the framework of rigorous Cohen-Lekner \cite{Cohen67} and Atrazhev \cite{Atrazhev85} theory. 
	In this theory, the electron transport through the liquid is considered as a sequence of single scatterings on the effective potential. Therefore, such a parameter as electron scattering cross section can be used in the liquid in a way similar to that of the gas \cite{Akimov21}. An important concept of the theory is the distinction between the energy transfer scattering, which changes the electron energy, and that of momentum transfer, which only changes the direction of the electron velocity. Both processes have been assigned separate cross sections \cite{Cohen67,Atrazhev85,Stewart10}: that of energy transfer (or else effective) and that of momentum transfer.  These are obvious analogs of those in the gas, namely of the total elastic and momentum transfer (transport) cross sections, respectively. The latest modifications of the theory can be found elsewhere \cite{Boyle15,Boyle16}.
	
	Accordingly, in this work the photon yields and spectra are calculated for NBrS EL in all noble liquids: in liquid He, Ne, Ar, Kr and Xe. The relevance of the results obtained to the development of noble liquid detectors for dark matter searches and neutrino detection is also discussed.

	\section{Theoretical formulas}
	
	To calculate the photon yields and spectra for NBrS EL in noble liquids we used the approach developed for noble gases in~\cite{Buzulutskov18}. Let us briefly recall the main points of this approach.

	The differential cross section for NBrS photon emission is expressed via electron-atom total elastic cross section ($\sigma _{el}(E)$)~\cite{Buzulutskov18,Park00,Firsov61,Kasyanov65,Dalgarno66,Biberman67}:
	\begin{eqnarray}
		\label{Eq-sigma-el} 
		\frac{d\sigma}{d\nu} = \frac{8}{3} \frac{r_e}{c} \frac{1}{h\nu} \left(\frac{E - h\nu}{E} \right)^{1/2} \times \hspace{40pt} \nonumber \\ \times \ [(E-h\nu) \ \sigma _{el}(E) \ + \ E \ \sigma _{el}(E - h\nu) ]  \; ,
	\end{eqnarray}
	where $r_e=e^2/m c^2$ is the classical electron radius, $c=\nu \lambda$ is the speed of light, $E$ is the initial electron energy and $h\nu$ is the photon energy. 
	
	To be able to compare results at different medium densities and temperatures, we need to calculate the reduced EL yield ($Y_{EL}/N$) as a function of the reduced electric field ($\mathcal{E}/N$), where $\mathcal{E}$ is the electric field and $N$ is the atomic density. The reduced EL yield is defined as the number of photons produced per unit drift path and per drifting electron, normalized to the atomic density; for NBrS EL it can be described by the following equation \cite{Buzulutskov18}: 
	\begin{eqnarray}
		\label{Eq-NBrS-el-yield} 
		\left( \frac{Y_{EL}}{N}\right)_{NBrS} =  \int\limits_{\lambda_1}^{\lambda_2}  \int\limits_{h\nu}^{\infty}\frac{\upsilon_e}{\upsilon_d} 
		\frac{d\sigma}{d\nu} \frac{d\nu}{d\lambda} f(E) \ dE \ d\lambda 
		\; , 
	\end{eqnarray}
	where $\upsilon_e=\sqrt{2E/m_e}$ is the electron velocity of chaotic motion,  $\upsilon_d$ is the electron drift velocity, $\lambda_1-\lambda_2$ is the sensitivity region of the photon detector,  
	$d\nu/d\lambda=-c/\lambda^2$, $f(E)$ is the electron energy distribution function normalized as
	\begin{eqnarray}
		\label{Eq-norm-f} 
		\int\limits_{0}^{\infty} f(E) \ dE = 1 \; .
	\end{eqnarray}
	The distribution functions with a prime, $f^\prime=f/E^{1/2}$, is often used instead of $f$, normalized as 
	\begin{eqnarray}
		\label{Eq-norm-fprime} 
		\int\limits_{0}^{\infty} E^{1/2} f^\prime(E) \ dE = 1 \; .
	\end{eqnarray}
	$f^\prime$ is considered to be more enlightening than $f$, since in the limit of zero electric field it tends to Maxwellian distribution.

	Consequently, the spectrum of the reduced EL yield is 
	\begin{eqnarray}
		\label{Eq-NBrS-el-yield-spectrum} 
		\frac{d (Y_{EL}/N)_{NBrS}}{d\lambda} = 
		\int\limits_{h\nu}^{\infty}\frac{\upsilon_e}{\upsilon_d} 
		\frac{d\sigma}{d\nu} \frac{d\nu}{d\lambda} f(E) \ dE \  
		\; . 
	\end{eqnarray}
	
	In our previous works \cite{Buzulutskov18,Borisova21}, the electron energy distribution function and drift velocity in noble gases, at a given reduced electric field, were calculated using Boltzmann equation solver \cite{Hagelaar05}. 
	
	In this work, we follow exactly the  Atrazhev paper \cite{Atrazhev85} to calculate the electron energy distribution function and drift velocity in noble liquids. Another modification is that the total elastic cross section in Eq.~\ref{Eq-sigma-el} is replaced with the energy transfer cross section for electron transport through the liquid. With these two modifications all the Eqs.~\ref{Eq-sigma-el},\ref{Eq-NBrS-el-yield},\ref{Eq-norm-f},\ref{Eq-NBrS-el-yield-spectrum} can directly apply to noble liquids.

	\section{Cross sections, electron energy distribution functions and drift velocities in noble liquids}
	
	According to Cohen-Lekner and Atrazhev theory the drift and heating of excess electrons by an external electric field in the liquid are determined by two parameters, the collision frequency of energy transfer ($\nu_{e}$) and that of momentum transfer ($\nu_{m}$)~\cite{Atrazhev85}:
	\begin{eqnarray}
		\label{Eq01} 
		\nu_{e} = \delta N \sigma_{e}(E)(2E/m)^{1/2} \: , \\
		\nu_{m} = N \sigma_{m}(E)(2E/m)^{1/2} \: , \\  
		\sigma_{m}(E) = \sigma_{e}(E)\widetilde{S}(E) \,. 
	\end{eqnarray}
	\noindent Here $N$ is the atomic density of the medium; $E$ is the electron energy; $\delta = 2m/M$ is twice the electron-atom mass ratio; $\sigma_{e}(E)$ and $\sigma_{m}(E)$ is the energy transfer (effective) and momentum transfer electron scattering cross section in the liquid, respectively; $\widetilde{S}(E)$ is the function, which takes into account liquid structure. 
	%Both collision frequencies will be used later to calculate the electron energy distribution function and drift velocity. 
	
\begin{figure}
		\includegraphics[width=0.99\columnwidth]{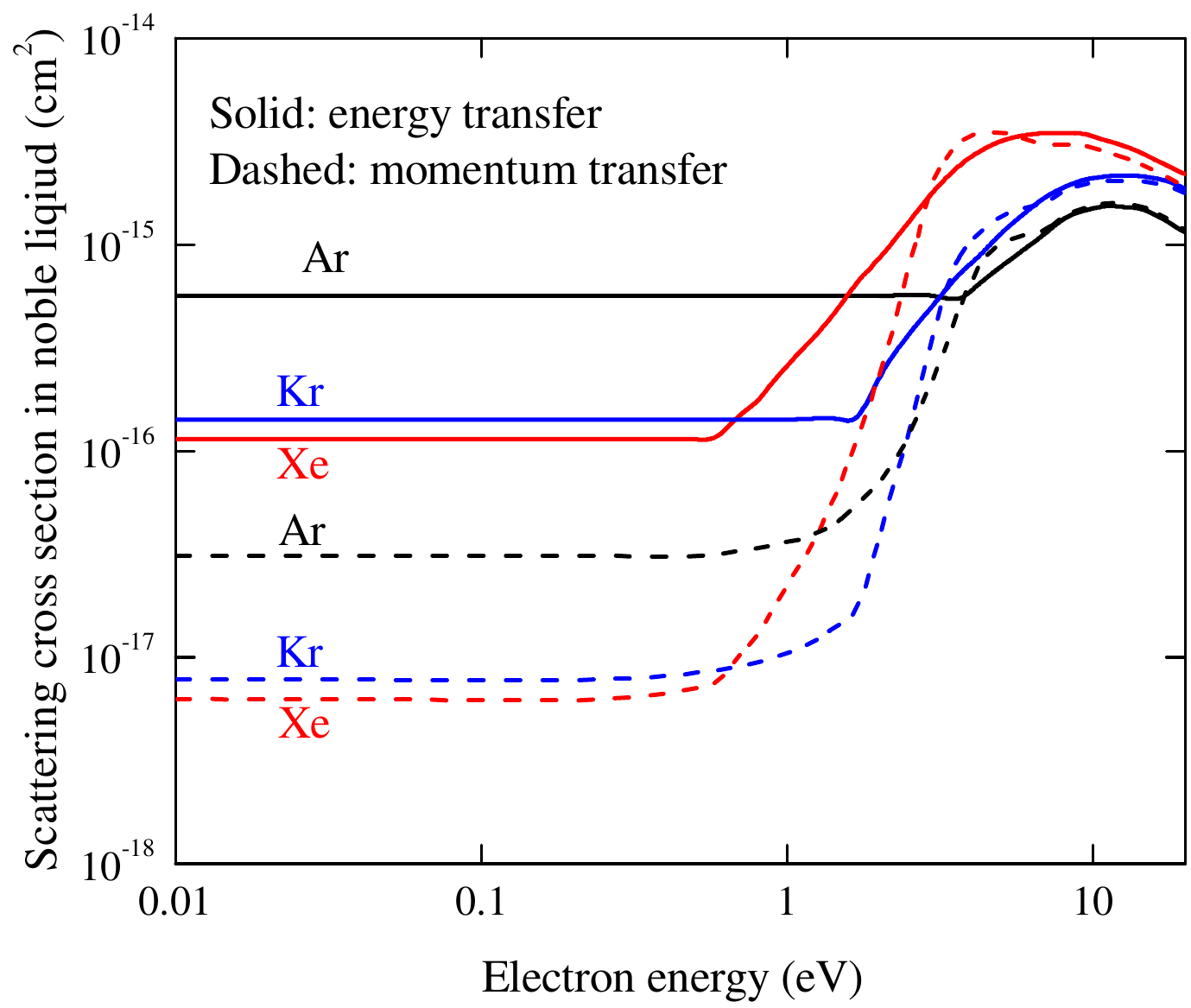}
		\includegraphics[width=0.99\columnwidth]{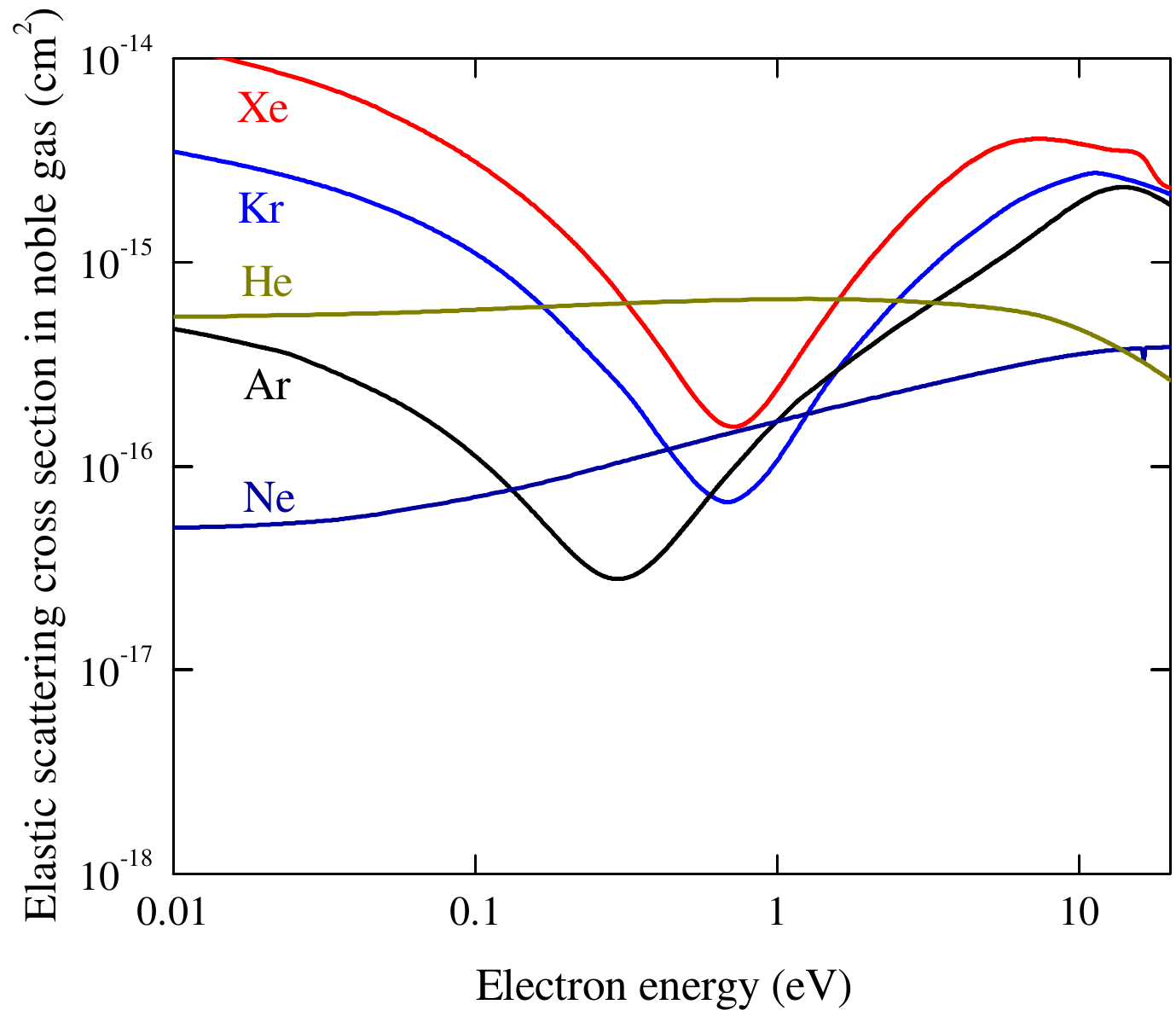}
		\caption{Top: Electron scattering cross sections in liquid Ar, Kr and Xe as a function of electron energy, namely that of energy transfer (or effective), $\sigma_{e}$, and that of momentum transfer, $\sigma_{m}$, both taken from~\cite{Atrazhev85}.
			Bottom: Electron scattering cross section in noble gases as a function of electron energy: that of total elastic for Ne, Ar, Kr, and Xe, taken from the BSR database~\cite{DBBSR}, and that of momentum transfer for He, taken from the Biagi database~\cite{DBBiagi}.}
		\label{fig01}
	\end{figure}

	To calculate collision frequencies one need to know $\sigma_{e}(E)$ and $\sigma_{m}(E)$; for liquid Ar, Kr and Xe these were given in \cite{Atrazhev85}: see Fig.~\ref{fig01} (top). 
	For comparison, Fig.~\ref{fig01} (bottom) presents the total elastic cross sections for gaseous Ne, Ar, Kr, and Xe taken from the BSR database~\cite{DBBSR}; since for He it is not available, the momentum transfer cross section is shown instead taken from the Biagi database~\cite{DBBiagi}. 
	
	The electron distribution function $f^\prime(E)$ in a strong electric field  is expressed via both collision frequencies \cite{Atrazhev85}:
	
	\begin{eqnarray}
		\label{Eq02} 
		f^\prime(E) = f(0) exp\left(-\int\limits_{0}^{E} \frac{3m\nu_{e}(E)\nu_{m}(E)}{2e^{2}\mathcal{E}^{2}}dE\right). 
	\end{eqnarray}
	The constant $f(0)$ is determined from the normalization condition 
	of Eq.~\ref{Eq-norm-fprime}.
	%$\int\limits_{0}^{\infty} E^{1/2} f^\prime(E) \ dE = 1$. 
	
\begin{figure}
		\includegraphics[width=0.99\columnwidth]{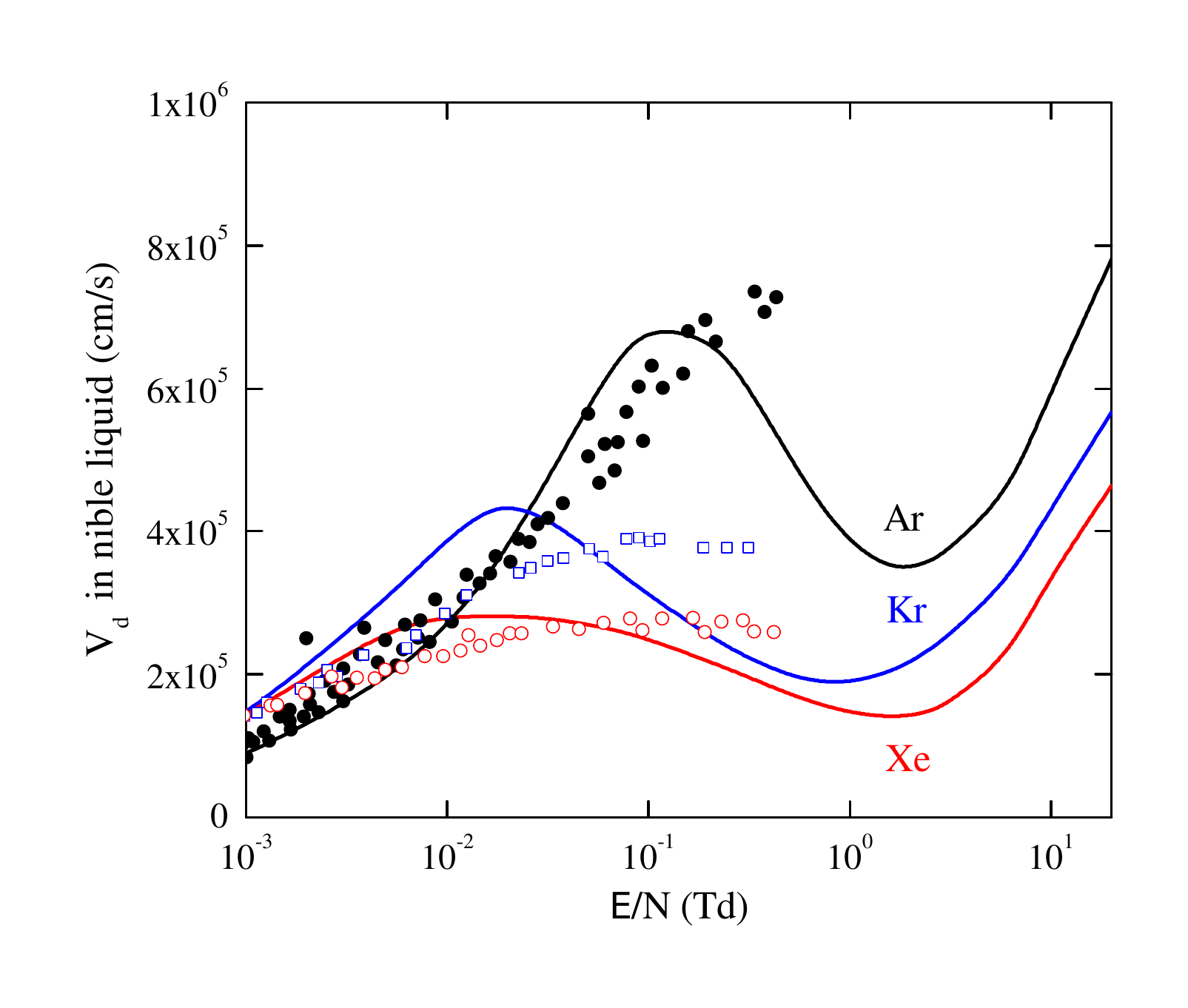}
		\caption{Comparison of electron drift velocity ($\upsilon_d$) in liquid Ar, Kr and Xe theoretically calculated in this work (curves) with that measured in experiment \cite{Miller68} (data points). The color of the curve and the data points is the same for a given noble liquid.}
		\label{fig02}
	\end{figure}

	Using the electron energy distribution functions, one can calculate the electron drift velocity in the liquid \cite{Atrazhev85}:
	\begin{eqnarray}
		\label{Eq03} 
		\upsilon_d = -\frac{2}{3}\frac{e\mathcal{E}}{m} \int\limits_{0}^{\infty} \frac{E^{3/2}}{\nu_{m}(E)} \frac{df^\prime}{dE} dE. 
	\end{eqnarray}
	It is shown in Fig.~\ref{fig02} as a function of the reduced electric field, the latter being expressed in Td units: 1~Td~=~$10^{-17}$~V~cm$^2$. It is possible to check the correctness of the distribution functions by comparing the calculated and measured electron drift velocities: this is done in Fig.~\ref{fig02} using the experimental data compiled in~\cite{Miller68}. It can be seen that the theoretical and experimental drift velocities are in a reasonable agreement, within a factor of 2, thus confirming the correctness of the calculated distribution functions for liquid Ar, Kr and Xe.
	
	It should be remarked that in light noble liquids, He and Ne, the Cohen-Lekner and Atrazhev theory cannot apply to calculate the electron energy distribution functions, since the appropriate cross sections for electron transport in the liquid, $\sigma_{e}(E)$ and $\sigma_{m}(E)$, are not available in the literature. Thereby in the following for these liquids a "compressed gas" approximation will be used, similarly to that developed in \cite{Borisova21}. In this approximation, the Eqs.~\ref{Eq-sigma-el},\ref{Eq-NBrS-el-yield},\ref{Eq-norm-f},\ref{Eq-NBrS-el-yield-spectrum} apply directly as for the gas, i.e. with electron energy distribution function and drift velocity obtained using Boltzmann equation solver, with the input elastic cross sections taken for the gas from Fig.~\ref{fig01} (bottom), and with the atomic density $N$ equal to that of the liquid. 

\onecolumn

\begin{table*} [h!]
	\caption{Properties of noble gases and liquids, and parameters of neutral bremsstrahlung (NBrS) electroluminescence (EL) theoretically calculated in this work.}
	\label{table}
	\begin{center}
		\begin{tabular}{p{0.5cm}p{6cm}p{1.5cm}p{1.5cm}p{1.54cm}p{1.5cm}p{1.5cm}}
			No & Parameter & He & Ne & Ar & Kr & Xe \\
			\\
			(1) & Boiling temperature at 1.0~atm, $T_b$~\cite{Fastovsky71} (K)  & $4.215$ & $27.07$ & $87.29$ & $119.80$ & $165.05$ \\
			%\\
			(2) & Gas atomic density at $T_b$ and 1.0 atm, derived from~\cite{Fastovsky71} (cm$^{-3}$) & $2.37\cdot10^{21}$ & $3.41\cdot10^{20}$ & $8.62\cdot10^{19}$ & $6.18\cdot10^{19}$ & $5.75\cdot10^{19}$ \\
			%\\
			(3) & Liquid atomic density at $T_b$ and 1.0 atm, derived from~\cite{Fastovsky71} and from ~\cite{Theeuwes70} for Xe (cm$^{-3}$) & $1.89\cdot10^{22}$ & $3.59\cdot10^{22}$ & $2.10\cdot10^{22}$ & $1.73\cdot10^{22}$ & $1.35\cdot10^{22}$ \\
			%\\
			%(4) & Reduced electric field corresponding to electric field in the liquid of 100 kV/cm (Td) & $0.53$ & $0.28$ & $0.48$ & $0.58$ & $0.74$\\
			%\\
			(4) & Threshold in electric field for excimer EL in noble liquid deduced from the corresponding threshold in noble gas by reduction to the atomic density of the liquid, obtained using data of \cite{Borisova21} (kV/cm) & $1134$ & $538$& $840$ & $519$ & $472$\\
			%\\
			(5) & Number of photons for NBrS EL in noble liquid produced by drifting electron in 1~mm thick EL gap at $T_b$ and 1.0~atm, at electric field of 100 kV/cm & $0.13$ & $2.5$& $0.93$ & $1.6$ & $1.1$\\
			%\\
			(6) & The same at 500 kV/cm & $4.3$ & $40$& $12$ & $24$ & $30$\\
			%\\
			%(6) & Threshold in reduced electric field for excimer EL in noble gas calculated in \cite{Borisova21} (Td) & $6.0$ & $1.5$& $4.0$ & $3.0$ & $3.5$\\
			
		\end{tabular}
	\end{center}
\end{table*}

\begin{multicols}{2}
	\twocolumn
	
	The values of the atomic densities for the gas and liquid phases at boiling temperatures at 1 atm are presented in Table~\ref{table}. We will see in the following on the example of heavy noble liquids that "compressed-gas" approximation works well: the difference in photon yields for NBrS EL between the "liquid" and "compressed-gas" approximation is not that large, remaining within a factor of 1.5.
	
	It should be also remarked, that all the calculations in this work were provided for atomic densities of the medium, liquid or gas, corresponding to the boiling temperature of a given noble element at 1 atm.  
	
	\section{Operational range of reduced electric fields in noble liquids for NBrS EL}
	
	It is obvious that NBrS EL in noble liquids is much weaker than excimer EL and thus becomes insignificant above the electric field threshold for excimer EL.  Table~\ref{table} gives an idea of these thresholds in noble liquid deduced from the corresponding threshold in noble gas by reduction to the atomic density of the liquid, obtained using the data of \cite{Borisova21}.  
	
	To compare with the results for noble gases, one also need to determine the operational range of reduced electric fields for NBrS EL in noble liquids from the experimental works, where it was presumably observed and where the operation electric field can be reliably estimated. Basically 3 works do fit to these conditions: that of \cite{Buzulutskov12}, operating the gas electron multiplier  (GEM,\cite{Sauli16}) in liquid Ar, that of \cite{Lightfoot09}, operating the thick GEM (THGEM, \cite{Breskin09}) in liquid Ar, and that of \cite{Aprile14}, operating the thin anode wire in liquid Xe. Deduced from the absolute electric field values given in \cite{Buzulutskov12} and \cite{Aprile14}, the required  range of reduced electric fields within which NBrS EL was presumably observed amounts to 0.1-5 Td. In particular, for liquid Ar this range corresponds to electric fields ranging from 21 to 1040 kV/cm. We will restrict our calculations to this range of fields.
	%to 0.28-0.30 Td \cite{Lightfoot09}, 0.24-0.67 Td \cite{Buzulutskov12} and 1-5 Td \cite{Aprile14}.
		
	\section{NBrS EL spectra and yields in noble liquids}
	
\begin{figure}
	\includegraphics[width=0.99\columnwidth]{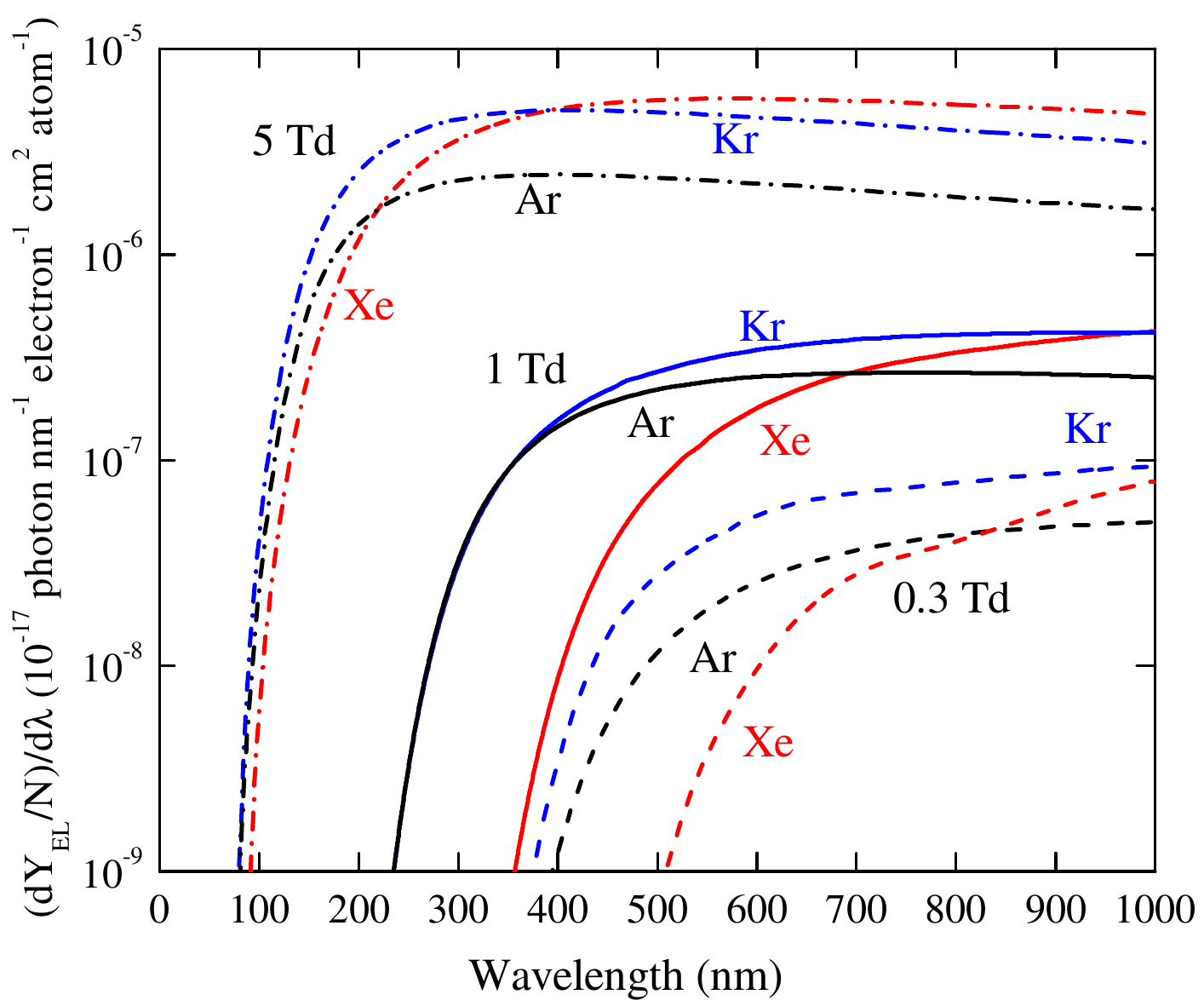}
	\caption{Spectra of the reduced EL yield for NBrS EL in liquid Ar, Kr and Xe at different reduced electric fields (0.3, 1 and 5 Td), calculated using Eq.~\ref{Eq-NBrS-el-yield-spectrum}.}
	\label{fig03}
\end{figure}

	Figs.~\ref{fig03} show the NBrS spectra of the reduced EL yield for liquid Ar, Kr and Xe at different reduced electric fields. The spectra were calculated by numerical integration of Eq.~\ref{Eq-NBrS-el-yield-spectrum}.
	
	One can see that NBrS EL spectra are similar in all noble liquids; moreover they look almost identical to those obtained in noble gases at the same reduced electric field: compare Fig.~\ref{fig03} to Fig.~10 of \cite{Borisova21} at 5 Td. The spectra are rather flat, extending from the UV to visible and NIR range at higher reduced electric field, e.g. at 5 Td. In each noble liquid, the NBrS EL spectrum has a broad maximum that gradually moves to longer wavelength with decreasing electric field. At lower reduced electric field,  in particular at 0.3 Td corresponding to 60 kV/cm in liquid Ar, the spectra have moved completely to the visible and NIR ranges. In all noble liquids, the spectra are mostly above 200 nm  (in the UV, visible and NIR range), i.e. just in the sensitivity region of commonly used photomultiplier tubes (PMTs) and silicon photomultipliers (SiPMs).
	
\end{multicols}

\twocolumn

\begin{figure}
	\includegraphics[width=0.99\columnwidth]{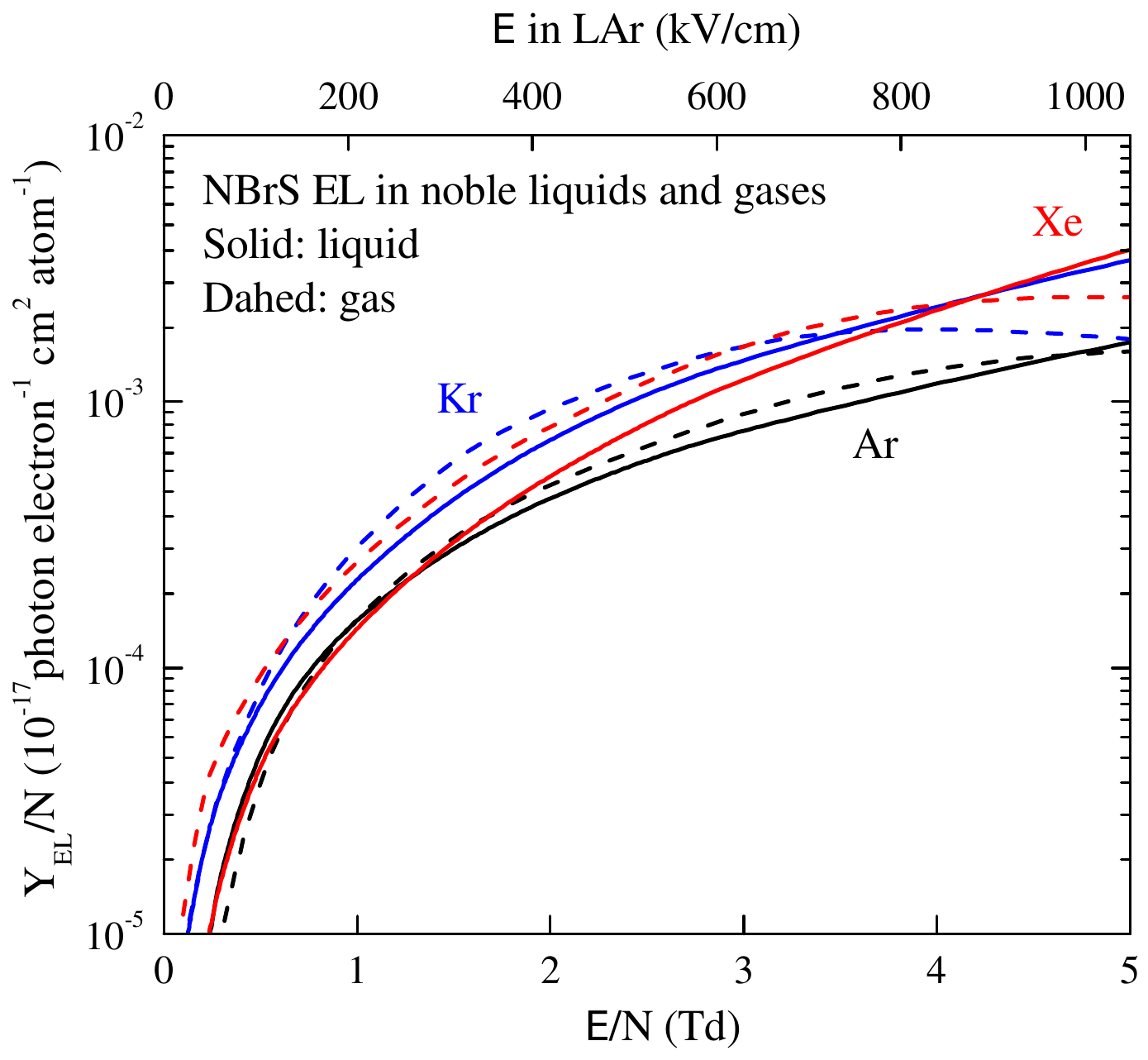}
	\caption{Reduced EL yield for NBrS EL at 0-1000 nm in liquid Ar, Kr and Xe as a function of the reduced electric field, calculated in this work in the framework of Cohen-Lekner and Atrazhev theory using Eq.~\ref{Eq-NBrS-el-yield} (solid lines).  For comparison, the reduced yield for NBrS EL at 0-1000 nm in noble gases is shown  calculated in \cite{Borisova21} using Boltzmann equation solver (dashed lines). The color of the curves are the same for a given noble element. The top scale shows the corresponding absolute electric field in liquid Ar.}
	\label{fig04}
\end{figure}

\begin{figure}
	\includegraphics[width=0.99\columnwidth]{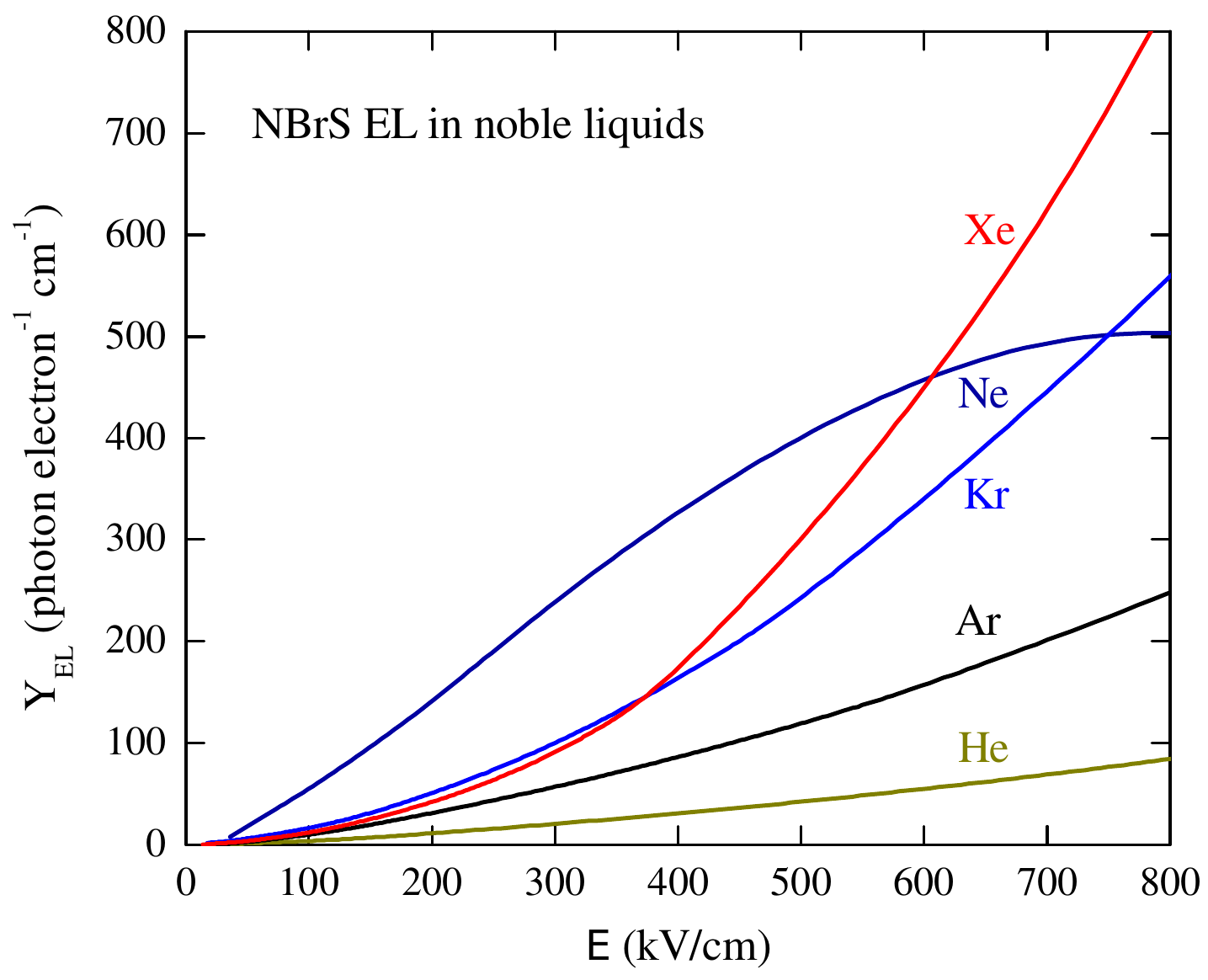}
	\caption{Absolute EL yield (number of photons per drifting electron per 1 cm) for NBrS EL at 0-1000 nm in noble liquids as a function of the absolute electric field, calculated in this work. For heavy noble liquids (Ar, Kr and Xe) the rigorous Cohen-Lekner and Atrazhev theory was used to calculate the electron energy and transport parameters in the liquid, while for light noble liquids (He and Ne) the "compressed gas" approximation was used.}
	\label{fig05}
\end{figure}

The EL yield for NBrS EL in noble liquids is presented in Figs.~\ref{fig04} obtained by numerical integration of Eq.~\ref{Eq-NBrS-el-yield}: the reduced EL yield is shown as a function of the reduced electric field. For comparison, the reduced yield for NBrS EL in noble gases is shown  calculated in \cite{Borisova21} using Boltzmann equation solver. 

Surprisingly, this "compressed-gas" approximation, successfully applied before to describe NBrS EL in noble gases, has led to almost the same results  as that of the rigorous "liquid" theory in terms of the reduced EL yields and spectra when formally extrapolated to the atomic density of the  noble liquid: for a given noble element and given reduced electric field the difference between them remains within a factor of 1.5 up to reduced electric field of 5 Td.

This fact indicates that the scaling law, stating that the reduced EL yield ($Y/N$) is a function of the reduced electric field ($\mathcal{E}/N$), is valid not only for noble gases, but also for noble liquids to some extent, at least as concerned the NBrS EL effect.
It also indicates on the applicability of the "compressed gas" approximation to noble liquids at moderate reduced electric fields, below 5 Td, thus justifying its use for light noble liquids, He and Ne, where the Cohen-Lekner and Atrazhev theory cannot be used due to the lack of the data.

Furthermore, Fig.~\ref{fig05} shows the practical photon yield   suitable for verifying in experimental conditions, namely the number of photons produced by drifting electron per 1 cm in all noble liquids, as a function of the absolute electric field. In this figure, for heavy noble liquids (Ar, Kr and Xe) the rigorous Cohen-Lekner and Atrazhev theory was used to calculate the electron energy and transport parameters in the liquid, while for light noble liquids (He and Ne) the "compressed gas" approximation was used, with the calculations identical to those of \cite{Borisova21}. The appropriate NBrS EL spectra and yields for He and Ne can be found in \cite{Borisova21}.

Table~\ref{table} (items 5 and 6) gives an idea of the magnitude of the NBrS EL effect in a practical parallel-plate EL gap, of a thickness of 1 mm:  at a field of 500 kV/cm the photon yield varies as 4, 40, 12, 24 and 30 photons for He, Ne, Ar, Kr and Xe, respectively. On the other hand,  at 100 kV/cm the photon yield is reduced by about an order of magnitude down to about 1 photon per drifting electron in almost all noble liquids. It is remarkable that up to 600 kV/cm, liquid Ne has the highest EL yield for NBrS EL, obviously due to much lower elastic cross section between 1 and 10 eV of the electron energy compared to other noble elements (see Fig.~\ref{fig01} (bottom)), resulting in stronger electron heating by the electric field and thus in more intense NBrS photon emission.

\section{Possible applications and discussion}

In order to produce noticeable NBrS EL in noble liquids, one should provide high enough electric fields, ranging from 50 to 500 kV/cm, in  practical devices. Based on previous experience, such devices might be GEMs \cite{Buzulutskov12}, THGEMs \cite{Lightfoot09} and thin anode wires \cite{Aprile14}. The parallel-plate EL gap of a thickness of 1 mm can also be considered, albeit being not tested in real experiment in noble liquids at such high fields. It should be remarked that the larger EL gap thickness, e.g. 1 cm, can hardly be used in practice due to the existing limit on high voltage breakdowns in noble liquids: the absolute voltage before breakdown cannot exceed values of about 100 kV in liquid He~\cite{Gerhold94} and several hundreds kV in other noble liquids \cite{Buzulutskov20,Auger16,Tvrznikova19}. 

It looks natural to use GEMs or THGEMs as EL plates instead of parallel-plate EL gaps in noble liquids, since the former are more resistant to breakdowns than the latter. Note that the NBrS EL spectrum is mostly in the visible and NIR range: see Fig.~\ref{fig03}. This implies a possible practical application of NBrS EL in noble liquid detectors, namely the method of direct optical readout of S2 signal in the visible range, i.e. without using a wavelength shifter (WLS). A similar technique has been recently demonstrated in two-phase Ar detector with direct SiPM-matrix readout using NBrS EL in the gas phase \cite{Aalseth21}. These results have lead us to the idea of using THGEM plates in combination with SiPM-matrices that have high sensitivity in the visible and NIR range, to optically record the S2 signal in single-phase noble liquid detectors for dark matter search and neutrino experiments. In addition, the recently proposed transparent very-thick GEM \cite{Kuzniak21} can be used as EL plate, with enhanced light collection efficiency.

We can verify the theory of NBrS EL in noble liquids by experiments where it was presumably observed, where the electric field is explicitly known and where it is known how to convert the emitted photons into recorded photoelectrons. At first glance, only two works qualify for these criteria:  \cite{Lightfoot09} and \cite{Aprile14}. 

In particular, in \cite{Lightfoot09} the operation electric field in the center of THGEM hole (1.5 mm height), of 60 kV/cm \cite{Buzulutskov12}, corresponds to $\mathcal{E}/N$=0.3~Td in liquid Ar, resulting in about 0.6 photons per drifting electron predicted by the NBrS EL theory according to Figs.~\ref{fig04} and \ref{fig05}. However, this is more than 2 orders of magnitude smaller than the light gain reported in \cite{Lightfoot09}. We therefore suggest to interpret the results of \cite{Lightfoot09} as caused by the presence of gas bubbles associated to THGEM holes, inside which proportional EL in the gas phase took place, similarly to what happens in Liquid Hole Multipliers \cite{Erdal20}.

\begin{figure}
	\includegraphics[width=0.99\columnwidth]{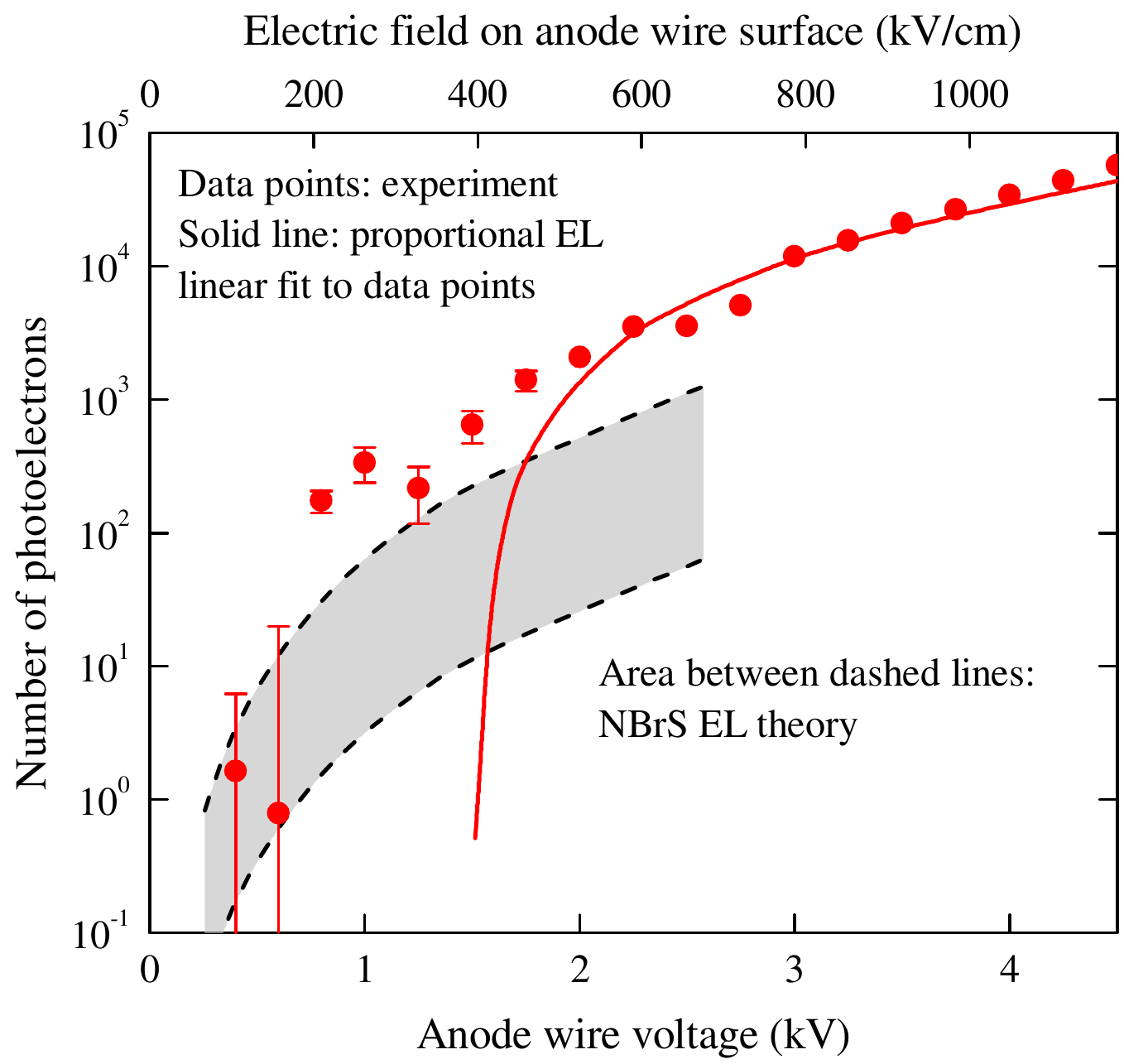}
	\caption{Number of photoelectrons recorded in liquid Xe by a PMT as a function of the voltage on $10~\mu m$ thick anode wire \cite{Aprile14}: the experimental data (data points) and linear fit of proportional EL to the data (solid line) are shown, the latter defining the threshold of excimer EL. Top scale shows the corresponding reduced electric field on anode wire surface. For comparison, the theoretical assessment of the number of photoelectrons due to NBrS EL obtained in this work is shown (area between dashed lines).}
	\label{fig06}
\end{figure}

In \cite{Aprile14}, where puzzling EL events were observed in liquid Xe under the threshold of excimer EL, the operation fields  near the anode wire were much higher, around 400 kV/cm. Fig.~\ref{fig06} shows the experimental data and linear fit of proportional EL to the data, the latter defining the threshold of excimer EL. In addition, the experimental conditions were explicitly described. This allowed us to predict the number of photoelectrons recorded by the PMT due to NBrS EL, although with some difficulties associated with highly inhomogeneous field near the wire. Due to the latter, Eq.~\ref{Eq-NBrS-el-yield} if applied directly gives only the lower limit of the event amplitude, since it does not take into account the electron diffusion, which significantly increases the travel time of the electron to the wire and thus the overall photon yield. We tried to take into account the diffusion effect: as a result, the theoretical prediction in Fig.~\ref{fig06} is shown in the form of an area between two dashed curves, thus setting the theoretical uncertainty. Within this uncertainty, the NBrS EL theory well describes the puzzling underthreshold events, namely their absolute amplitudes and the dependence on the anode voltage, which might be treated as the first experimental evidence for NBrS EL in noble liquids.

\section{Conclusion}

In this work we systematically studied the effect of neutral bremsstrahlung (NBrS) electroluminescence (EL) in all noble liquids:  the photon yields and spectra for NBrS EL have for the first time been theoretically calculated in liquid He, Ne, Ar, Kr and Xe. For heavy noble liquids, the calculations were done in the framework of Cohen-Lekner and Atrazhev theory describing the electron energy and transport parameters in the liquid medium. 

Surprisingly, the "compressed-gas" approximation, successfully applied before to describe NBrS EL in noble gases, has led to almost the same results  as that of the rigorous "liquid" theory in terms of the reduced EL yields and spectra when formally extrapolated to the atomic density of the  noble liquid. 

The predicted magnitude of the NBrS EL effect in a practical parallel-plate EL gap, of a thickness of 1 mm, is noticeable: at a field of 500 kV/cm the photon yield varies from 12 to 30 and 40 photons per drifting electron in liquid Ar, Xe and Ne respectively. The NBrS EL spectra  in noble liquids are in the visible and NIR range.
%In particular, the effect of NBrS EL in liquid Xe could explain puzzling  EL events under the threshold of excimer EL observed earlier in \cite{Aprile14}. 

The practical applications of the results obtained might be the use of THGEMs as EL plates in combination with SiPM-matrices, to optically record the S2 signal in single-phase noble liquid detectors for dark matter search and neutrino experiments.

\acknowledgments
This work was supported by Russian Science Foundation (project no. 19-12-00008). It was done within the R\&D program of the DarkSide-20k experiment. 

\bibliographystyle{eplbib}
\bibliography{mybibliography}   % name your BibTeX data base

\end{document}